\documentstyle[epsf,rotate]{mn}

\title[Non-barred boxy profiles]{Edge-on boxy profiles in non-barred disc
galaxies}

\author[P.A. Patsis et al]{P.A.~Patsis,$^1$ E.~Athanassoula,$^2$
   P.~Grosb{\o}l,$^3$ Ch.~Skokos$^{1,4}$\\ 
$^1$ Research
Center of Astronomy, Academy of Athens, Anagnostopoulou 14, GR-10673
Athens, Greece\\ 
$^2$ Observatoire de Marseille, 2 Place Le Verrier,
F-13248 Marseille Cedex 4, France\\
$^3$ ESO, Karl Schwarzschild Str.2, D-85748 Garching bei
               M\"unchen, Germany\\
$^4$ Division of Applied Analysis, Department of Mathematics and Center for
Research and Application of Nonlinear Systems (CRANS),\\ University of Patras,
GR-26500, Patras, Greece
} 

\date{Accepted ....  Received ....; in original form ....}
\pagerange{\pageref{firstpage}--\pageref{lastpage}}
\pubyear{2001}

\begin{document}

\maketitle   
\label{firstpage}  

\begin{abstract}
  Boxy edge-on profiles can be accounted for not only in models of
  barred galaxies, but also in models of normal (non-barred)
  galaxies. Thus, the presence of a bar is not a sine
  qua non condition for the appearance of this feature, as often assumed.
  We show that a `boxy' or a `peanut' structure in the central
  parts of a model is due to the presence of vertical resonances at
  which stable families of periodic orbits bifurcate
  from the planar x1 family. The orbits of these families reach in their
  projections on the equatorial plane a maximum distance from the center,
  beyond which they increase their mean radii by increasing only their
  deviations from the equatorial plane.  The resulting orbital
  profiles are `stair-type' and constitute the backbone for the observed
  boxy structures in edge-on views of $N$-body models and, we believe, in
  edge-on views of disc galaxies. Since the existence of vertical resonances is
  independent of barred or spiral perturbations in the disc, `boxy' profiles
  may appear also in almost axisymmetric cases.
\end{abstract} 

\section{introduction}
A significant number of edge-on disc galaxies display in their inner regions a
`boxy' or `peanut'-shaped (hereafter b/p) structure. It is believed
(L\"{u}tticke, Dettmar \& Pohlen 2000), that more than 45\% of disc galaxies
have this kind of edge-on profile. Typical examples are NGC~2424, NGC 6771,
NGC~5746, IC~4767, Hickson 87a and the Milky Way. Several authors (Combes \&
Sanders 1981; Pfenniger 1984, 1985; Combes, Debbasch, Friedli et al. 1990,
Pfenniger \& Friedli 1991, Raha, Sellwood J., James et al. 1991, Kuijken \&
Merrifield 1995, Bureau \& Freeman 1999) have related these profiles to the
presence of a strong bar. The tangential force in these bars is typically of
the order of 25\% of the axisymmetric one (Combes \& Sanders 1981).

Patsis \& Grosb{\o}l (1996) have shown that b/p orbital profiles appear also
in cases with a spiral instead of a bar perturbation. Recently Athanassoula
(unpublished) made a large number of $N$-body simulations to follow
the formation and evolution of bars in isolated
bar-unstable discs. Some of the models show in their edge-on views a
conspicuous b/p morphology, while their face-on views show
clearly that they are non-barred, and even in some cases almost
axisymmetric. It is clear that, at least in these cases, the b/p morphology is
due to the internal dynamics of the self-consistent model and not to merging 
phenomena.

In this paper we will first describe a particularly illustrative
simulation (section 2). We then use orbital theory to understand the
dynamics of b/p structures. We use a purely axisymmetric potential
consisting of a disc and a halo component (section 3), in order to
identify the orbits that constitute the backbone of the b/p structure
(section 4). The edge-on profiles are discussed in section 4. We
discuss our results in section 6 
and we stress the fact that it is the existence of vertical resonances
per se and not the kind of perturbation that gives rise to the appearance of
b/p morphologies.

\section{The $N$-body model}
Athanassoula's $N$-body stellar model starts with initial conditions created
by the method of Hernquist (1993) and consists of an exponential disc with a
$sech^{2}$ vertical dependence, and a halo profile proposed by Hernquist
(1993). The detailed description of the disc and halo initial density
distributions can be found in Athanassoula \& Misiriotis (2002; equations (1)
and (3)).

In computer units the mass of the disc is taken $M_d =1$, the halo mass $M_h
=5$, the disc scale length $h = 1$, the scale height of the disc $z_0 =
0.2$ and the halo scale length $\gamma =5$. The rest of the parameters of the
model in equations (1) and (3) of Athanassoula \& Misiriotis (2002) are as
described in Section 2 of the above mentioned paper.  In the simulation we
present here we can assume the unit of mass to be $5 \times 10^{10} M_{\sun}$,
the length unit 3.5~kpc, the unit of velocity 248~km s$^{-1}$ and the time
unit $1.4 \times 10^{7}$ yr.  The simulation has 200000 particles in the disc
and the halo is live and composed of 931206 particles. The Toomre
parameter Q is 1.8 (Toomre 1964).  We
underline the lack of an explicit bulge component.  The simulation was
carried out on a Marseille 
Observatory GRAPE-5 system using a tree-code similar to the one described in
Athanassoula, Bosma, Lambert et al.  (1998). For more details we again
refer the reader to Athanassoula \& Misiriotis (2002). 
\begin{figure}
\begin{center}
\epsfxsize=6.5cm \epsfbox{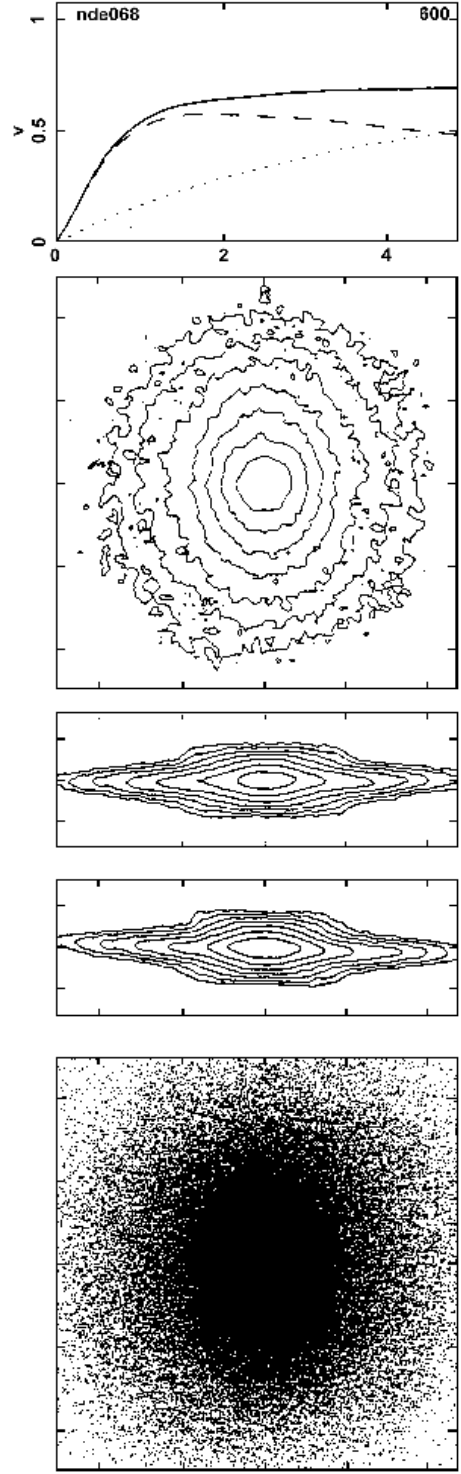}
\end{center}
\vspace{-0.5cm}
\caption{ Basic information on the stellar $N$-body simulation. From top to
  bottom are depicted the circular velocity curve (the dashed line gives the
  disc and the dotted the halo contribution), the isodensities of the disc
  particles projected face-on, side-on and end-on and finally the dot-plot of
  the particles in the ($x,y$) plane. The side of the box for the face-on
  views is 10 units, i.e. - with the adopted normalisation - 35~kpc, and the
  height of the boxes for the edge-on 
  views 3.33 units, i.e. $\approx$ 11.65~kpc.  In the top panel we give the
  name of the model and the time at which the snapshot is taken.}
\label{lia}
\end{figure}
A characteristic snapshot of the simulation after 8.4 Gyr from the start is
given in Fig.~\ref{lia}. The upper panel gives the circular velocity curve,
which is rising and reaches approximately 0.68 (i.e. $\approx$ 170 km
s$^{-1}$) at 5 unit lengths (i.e. 17.5 kpc). The dashed and dotted lines give
the disc and the halo contribution, respectively.  The second, third and
fourth panel from the top give the isodensities of the disc particles
projected face-on, side-on and end-on respectively. The lower panel gives the
dot-plot of the particles in the ($x,y$) plane. The side of the box for the
face-on views is 10 units (i.e. 35~kpc), so the height of the boxes for the
edge-on views 3.33 units (i.e. $\approx$ 11.65~kpc). From the face-on view it
is evident that the model essentially does not have a bar. One can speak about
a weak overall oval distortion. Such a weak perturbation can be detected in a
large fraction of disc galaxies as an m=2 component of low amplitude.
Nevertheless, both side-on and end-on views are obviously boxy. The fact that
the relative extent of the b/p feature is about the same when projected on the
horizontal axes reflects the fact that the isocontours of the density in the
face-on view are nearly round.

\section{The orbital model}
 
For our orbital calculations we will use a general axisymmetric
potential consisting of a disk and a halo component. We adopt a
Miyamoto and Nagai (1975) disc potential, $\Phi_{D}$, which 
in cylindrical coordinates has the form:

\begin{equation}
\Phi_{D}(r,z)= - \frac{G\;M_{D}}{\sqrt{r^{2}+[a+\sqrt{z^{2}+b^{2}}]^{2}}}
\end{equation}

\begin{figure}
\rotate[r]{ 
\epsfxsize=6cm \epsfbox{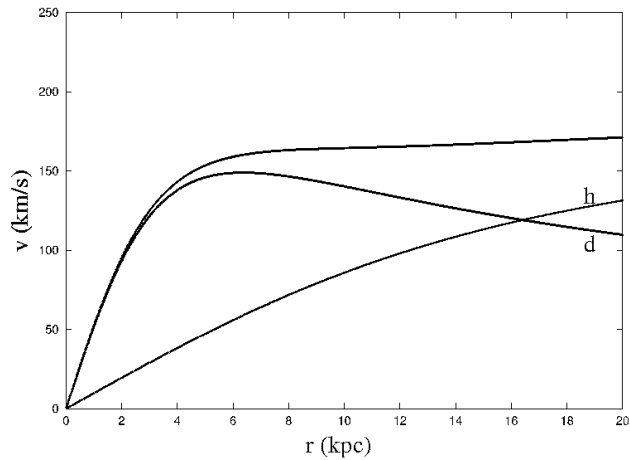}}
\caption{The rotation curve for the orbital model. It shows that the mass
distribution in the $N$-body and the orbital model is very close. The disc
contribution is indicated with a `d' and that of the halo with an `h'.}
\label{rotcur}
\end{figure}

In the above the parameter $M_{D}$ refers to the disc
mass, $a$ and $b$ are the horizontal and vertical scale lengths
respectively, and $r$ and $z$ are the cylindrical coordinates. The
halo potential, $\Phi_{H}$, is given by

\begin{equation}
\Phi_{H}(r,z)=\frac{v_{H}^{2}}{2}\;\ln\left(
1+\frac{1}{r_{c}^{2}}\,(r^{2} + z^{2})\right)
\end{equation}
where $v_{H}$ is the limiting circular velocity as $r \rightarrow
\infty$, and $r_{c}$ is the core radius of the halo. 
The total potential used for the orbital calculations is of the form:
$\Phi = \Phi_{D} + \Phi_{H}$, and the adopted values of the
parameters are  
$ M_{D}= 6\times~10^{10}~M_{\sun}$, 
$a= 3~$kpc,
$b=1.5~$kpc,
$r_{c}=18~$kpc and
$v_{H}=176.8~$km s$^{-1}$.
The rotation curve (Fig.~\ref{rotcur}) reproduces fairly well that of
the $N$-body model presented in the previous section.
One can clearly see that we have a model with a maximum disc, as in the 
case of the $N$-body model. We remind that 5 length units in the $N$-body
model correspond to 17.5~kpc, and its velocity unit is 248 km s$^{-1}$. 

\section{Orbits and orbital behavior}
The calculations have been done in cartesian coordinates in a frame of
reference rotating
around the $z$-axis. Thus, the
Hamiltonian governing the motion of the test particles is:
\begin{equation}
H=1/2(\dot{x}^{2}+\dot{y}^{2}+\dot{z}^{2})+\Phi(x,y,z)-1/2\,\Omega_{p}^{2}
(x^{2}+y^{2}),
\end{equation}
where $\Omega_{p} $ is the pattern speed. In the following we will
denote by $E_j$ the numerical value of $H$ and adopt for the pattern
speed the value $\Omega_{p}=11.9$ km s$^{-1}$ kpc$^{-1}$, which
places corotation approximately at 14~kpc.

On the equatorial plane we can define the radial resonances between the
epicyclic frequency and the angular velocity in the rotating frame. In a 3D
model we can also define vertical resonances, which involve the vertical
instead of the epicyclic frequency (for definitions see e.g. Binney
\& Tremaine (1987). The
first vertical resonance in our model is the 3:1. 

The main family in our model is the family of direct circular periodic
orbits on the $z$=0 plane. These orbits, in the
presence of a spiral or barred perturbation become ellipses,
the well known x1 orbits (see e.g. Contopoulos and Grosb{\o}l 1986,
1989). By analogy, we will call this family x1 also in our
axisymmetric model. In 2D models the orbits of this family support the
spiral or bar structure.  The stability of a 
periodic orbit is characterized by the behaviour of two indices $b_{1}$ and
$b_{2}$. In this study, the stability index $b_{1}$ is associated to the
motion perpendicular to the equatorial plane, while $b_{2}$ is associated to
radial perturbations. A family is stable if both stability indices $b_{i}$ are
$-2<b_{i}<2$ (Hadjidemetriou 1975). For more details on the stability of
families of periodic orbits in 3D systems the reader may refer to Contopoulos
and Magnenat (1985).

\begin{figure}
\rotate[r]{ 
\epsfxsize=6.0cm \epsfbox{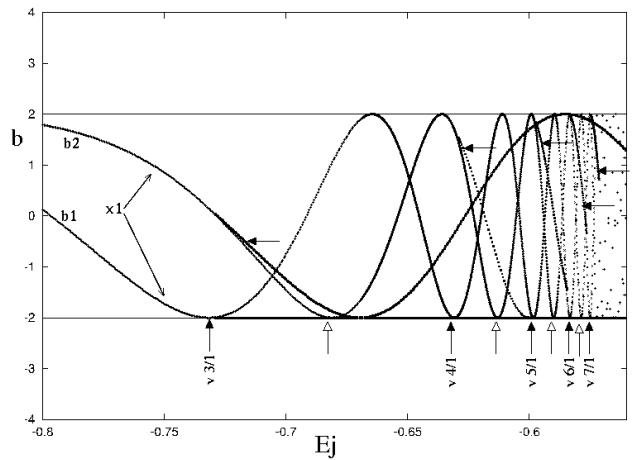}}
\caption{
The stability indices $b_{1}$ and $b_{2}$ of the central family x1 as a
function of $E_j$, where $E_j$ is the value of the Hamiltonian. 
We give also the stability indices of the families 
bifurcated at the vertical resonances 3:1, 4:1, 5:1, 6:1 and 7:1.
One of their stability indices remains always equal to 2, while the other one
oscillates between $-2$ and 2.}
\label{stabi}
\end{figure}

The most important families of periodic orbits for the dynamics of a 3D disc
galaxy are the central family and those bifurcated from it at the vertical
$n:1$ resonances, where $n$ is a small integer. At these
resonances, in the axisymmetric case, index $b_{1}$ of family x1 becomes
tangent to the $b=-2$ axis. The bifurcated families come actually in
pairs\footnote{The two families, which bifurcate at a tangency of a
  stability index with the $-2$ axis in the axisymmetric case, have a
  stable and an unstable 
  counterpart in the non-axisymmetric case} and are in this case marginally
stable because they always have one of their stability indices on the $b=-2$
axis, while the other remains always between $-2$ and $2$. The variation of
the stability indices $b_{1}$ and $b_{2}$ of the x1 family, as well as
those of the
families bifurcated at the vertical resonances 3:1, 4:1, 5:1, 6:1 and 7:1 are
given in Fig.~\ref{stabi} as a function of the Jacobian $E_j$. The vertical
black arrows indicate the points where index $b_{1}$ of x1 becomes tangent to
the $b=-2$ axis at the vertical resonances. Horizontal arrows point to the
index of each bifurcating family which oscillates between $-2$ and 2. They are
given always close to the value of the Hamiltonian at which the family
bifurcates from the x1.
The introduction of a barred or spiral perturbation in the model brings in the 
system a stable and an unstable family. For the stable one the 
index which in the axisymmetric case was lying on the $b=-2$ axis
now becomes absolutely smaller than 2, so the 
family remains stable over a large radial region (Patsis \& Grosb{\o}l 1996;
Skokos, Patsis, Athanassoula 2002a,b). The vertical white arrows indicate the
tangencies of the index $b_{2}$ of the x1 family with the $b=-2$ axis
at the radial resonances. 

\section{The edge-on profile}
Most stars in a galaxy will move along non-periodic orbits trapped around
stable periodic orbits (Poincar\'{e} 1892). Thus, the topology of the main
families of periodic orbits will determine the basic features in the galaxy.
In the present case the main families are the central family and the families
bifurcated at the vertical resonances, which exist as two branches symmetric
with respect to the equatorial ($z=0$) plane. The response density
profile is a weighted average of the individual orbit contributions. As a
weight, we use the disc density, $\rho_D(r,z)$, where $\rho_{D}$ is the density
corresponding to the Miyamoto disc calculated at the position $(r,z)$,
where $r$ is the radius of the circular
equatorial-plane orbit which has the same $E_j$ as the orbit to be weighted
and $z$ is the mean vertical distance of the orbit
from the equatorial plane. We have adopted the density of the luminous
(disc) component because the orbital profiles will have to reproduce
the light distribution, when we compare to real galaxies, or the disc
component of the $N$-body simulations.

Since our model is axisymmetric
we can use arbitrarily any two orthogonal axes on the equatorial plane in
order to calculate periodic orbits. This means that a periodic orbit
rotated around the axis of symmetry ($z$-axis) is also a periodic
orbit of the system. The projections of 
these periodic orbits on a given axis will always be confined within 
certain limits determined by a maximum length and a maximum height.
We also note that the orbits of the two families that bifurcate 
from the central family at the tangencies of the $b_1$ index with the $b=-2$ axis 
(Fig.~\ref{stabi}), at a given $E_j$ value, are topologically similar.

\begin{figure}
\epsfxsize=8cm \epsfbox{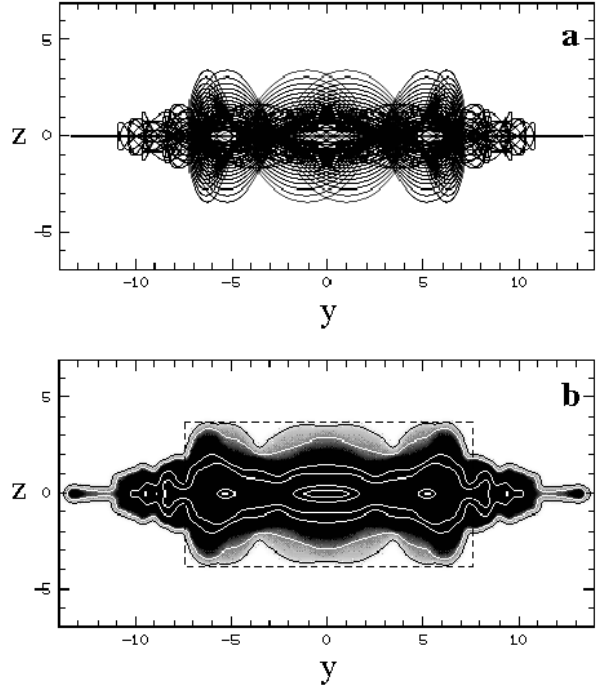}
\caption{The $(y,z)$ orbital profile. In (a) the profile of the
  weighted orbits, and in (b) the corresponding blurred image. The dashed box
  indicates the area which is filled if all orbits at a given energy are considered. }
\label{yz}
\end{figure}

We have calculated at each energy ($E_j$) only one periodic orbit per family 
along our $x$-axis. We started with the orbit which has the same radius with the
circular orbit at the $E_j$ where the family is bifurcated, and we followed
the evolution of the family by finding the orbits along the specific $x$-axis.
These orbits have been used for constructing the profiles in
Fig.~\ref{yz}. Nevertheless, at each energy, one can find, by
rotation, an infinite number of
representatives of the same family, due to the axisymmetric nature of the
potential. It is as viewing a given orbit from all possible view angles by
rotating our point of view around the $z$-axis, while staying
always on the equatorial plane. 

In the $(y,z)$ profile, created by (marginally) stable orbits of the families
considered in Fig.~\ref{stabi} (Fig.~\ref{yz}a), we observe a `stair-type'
structure with a central boxy region. This central region is formed by the
orbits of the two branches (symmetric with respect to the equatorial
plane) of the 3D family bifurcated at the vertical 3:1 
resonance. By applying a smoothing filter on the image with the weighted
orbits we obtain a blurred profile, which, to a first approximation, can be
considered as the profile for the density distribution of the model. This
blurred profile is shown in Fig.~\ref{yz}b.
By considering all possible orbits at each energy, the area inside the
rectangle drawn with a dashed line in Fig.~\ref{yz}b will be filled.
The $(x,z)$ profile, when considering all orbits, will be identical to
the $(y,z)$ one, as expected. 
In Fig.~\ref{yz}b we illustrate the resulting morphology by using isocontours.
Clearly the profiles of the orbital model (Fig.~\ref{yz}) and
the profiles of the $N$-body simulation (Fig.~\ref{lia}) have many
similarities. First of all we have a boxy structure confined in
radii less than 2 length units of the $N$-body simulation, which correspond to
7~kpc. This is practically the radius in Fig.~\ref{yz}  inside which we
find the boxy structure of the orbital model. Also the `stair-type' edge-on
profiles of the weighted orbits is in good agreement with the outer parts of
the edge-on profiles of the $N$-body model. 

\section{Discussion}
In the present study we propose a mechanism, which can account for
boxy structures in the edge-on profiles of non-barred disc
galaxies. In the example we present here, it is the presence of the
vertical 3:1 
resonance which introduces this structure in the system. It is, however, 
not necessary to have a particular vertical 
resonance in order to form a b/p profile. Furthermore, since radial
and vertical resonances can be defined even in an
axisymmetric case, one can find even in the axisymmetric model the bifurcating
orbits that give rise to these features. In fact, the only two necessary
ingredients are: 

- the presence of a vertical $n:1$ resonance, where $n$ a small
  integer, so that a new family, bifurcating at this point, is
  introduced in the system

- the bifurcated family should be stable over a sufficiently large
$E_j$ interval and should trap a sufficient number of regular orbits
around it.

If these two conditions are fulfilled, the `stair-type' orbital profile
follows naturally, because the successive 3D families lower their mean heights
as the energy at which they are bifurcated from x1 increases.  In addition, as
energy increases, the successive orbits of a bifurcating 3D family increase
their mean spherical radius by growing more in $z$ than in their cylindrical
radius and, beyond a critical energy, by growing practically only in the
vertical direction.  Since their cylindrical radius -- or extent along the
equatorial plane -- is thus limited, while their vertical extent is large, at
least for large values of the Jacobi energy $E_J$, they contribute to a boxy
profile. As we have shown here, it is not necessary to have a strong
perturbation in order for the vertical extent to be important. Thus the boxy
feature can be strong and clearly defined, even in a purely axisymmetric
model.

If we introduce a perturbation, the topology of the relevant orbital
families remains the same and
they will be stable rather than marginally stable (Patsis \&
Grosb{\o}l 1996; Skokos et al. 2002a,b). There will, however, be one
morphological difference, due to the fact that in the non-axisymmetric
case the rotational symmetry is broken. In this case it will not be
possible to have orbits of any desired azimuthal orientation, and as a
result we will have 
a `peanut-shaped' profile, at least for a range of viewing
angles, instead of a boxy one. 

Lack of a boxy structure in a model indicates one of the following
three possibilities: Either  
vertical resonances with small $n$:1 do not exist in the system. Or the family
bifurcated at the vertical resonance with the lowest $E_j$ value has
too large unstable parts. Or, for reasons that could be linked to the
formation history of the galaxy, too few stars are on orbits trapped
around the stable periodic $n$:1 orbits. The
conditions needed for building a b/p profile may be favored by the presence of
the bar, but the bar per se is not the reason that edge-on disc galaxies have
boxy profiles. Of course our mechanism relies on the existence of a
pattern speed which does not vary much with time. In other words it
implicitly assumes the existence, or past existence, of some
non-axisymmetric feature, albeit of perhaps infinitesimal amplitude.

Bureau \& Athanassoula (1999) and Athanassoula \& Bureau (1999) developed
diagnostics to detect the presence and orientation of a bar in edge-on disc
galaxies. They detected in most of the peanut shaped edge-on galaxies in
Bureau \& Freeman (1999) the signature of a `x2-flow', in the
position-velocity diagrams. This was taken as the manifestation of the
presence of a bar. However, in a few cases, this feature was
absent. Athanassoula \& Bureau accounted the lack of such a feature
either to a lack of an ILR or to a lack of emitting gas around the ILR
region. The present study adds a third possibility, namely that the
galaxy is not barred.

\section*{Acknowledgments}
We acknowledge fruitful discussions and very useful comments by Prof.
G.~Contopoulos. We also thank the anonymous referee for her/his comments which
improved the paper. This work has been supported by the Research Committee of
the Academy of Athens. E. Athanassoula would also like to thank the IGRAP, the
Region PACA, the INSU/CNRS and the University of Aix-Marseille I for funds to
develop the GRAPE computing facilities used for the simulations discussed in
this paper.  P.A.~Patsis and Ch.~Skokos thank the Laboratoire d'Astrophysique
de Marseille, for an invitation. Ch.~Skokos was supported by the
``Karatheodory'' fellowship No 2794 of the University of Patras.

\bsp

\label{lastpage}     
\end{document}